\documentclass[preprintnumbers, showkeys,floatfix,12pt,prd,onecolumn, superscriptaddress,nofootinbib]{revtex4-2}
\usepackage{bm}
\usepackage{comment}

\usepackage{mathtools}
\usepackage{amsfonts}
\usepackage{latexsym}
\usepackage[latin1]{inputenc}
\usepackage{graphicx}
\usepackage{amsmath}
\usepackage{palatino}
\usepackage{mathpazo}
\usepackage{textcomp}
\usepackage{float}
\usepackage{booktabs}
\usepackage{dcolumn} 
\usepackage{caption}
\usepackage{subcaption}
\usepackage{ragged2e}
\usepackage{hyperref}
\usepackage{amsmath}
\usepackage{xcolor}
\usepackage{textcomp}
\usepackage{orcidlink}
\usepackage{epsfig}
\usepackage{commath}
\usepackage{multirow}
\usepackage{algorithm}
\usepackage{algpseudocode}
\usepackage{array}
\hypersetup{colorlinks,citecolor=blue}
\hypersetup{colorlinks=true,linkcolor=red,filecolor=magenta,    urlcolor=blue}

\begin{document}
\begin{center}
\hrule height 4.1pt
\vspace{0.5cm}
{\Large\textbf{Enhancing Quantum Support Vector Machines through Variational Kernel Training}} 
\vspace{0.5cm}
\hrule height 1.2pt
\end{center}
\author{N. INNAN\orcidlink{0000-0002-1014-3457}}
\email{nouhailainnan@gmail.com}

\begin{abstract}
Quantum machine learning (QML) has witnessed immense progress recently, with quantum support vector machines (QSVMs) emerging as a promising model. This paper focuses on the two existing QSVM methods: quantum kernel SVM (QK-SVM) and quantum variational SVM (QV-SVM). While both have yielded impressive results, we present a novel approach that synergizes the strengths of QK-SVM and QV-SVM to enhance accuracy. Our proposed model, quantum variational kernel SVM (QVK-SVM), leverages the quantum kernel and quantum variational algorithm. We conducted extensive experiments on the Iris dataset and observed that QVK-SVM outperforms both existing models in terms of accuracy, loss, and confusion matrix indicators. Our results demonstrate that QVK-SVM holds tremendous potential as a reliable and transformative tool for QML applications. Hence, we recommend its adoption in future QML research endeavors.

\end{abstract}

\date{\today }

\color{black} 


\affiliation{Quantum Physics and Magnetism Team, LPMC, Faculty of Sciences Ben M'sick, Hassan II University of Casablanca, Morocco.}
\affiliation{Quantum Formalism Fellow, Zaiku Group Ltd, Liverpool, United Kingdom.}

\author{MAZ. KHAN\orcidlink{0000-0002-1147-7782}}
\email{muhammadalzafark@gmail.com}
\affiliation{Robotics, Autonomous Intelligence, and Learning Laboratory (RAIL), School of Computer Science and Applied Mathematics, University of the Witwatersrand, 1 Jan Smuts Ave, Braamfontein, Johannesburg 2000, Gauteng, South Africa}
\affiliation{Quantum Formalism Fellow, Zaiku Group Ltd, Liverpool, United Kingdom.}

\author{B. PANDA}
\affiliation{Indian Institute of Science Education and Research (IISER), Berhampur, Odisha, India.} 

\author{M. BENNAI\orcidlink{0000-0002-7364-5171}}
\affiliation{Quantum Physics and Magnetism Team, LPMC, Faculty of Sciences Ben M'sick, Hassan II University of Casablanca, Morocco.} 
\affiliation{Lab of High Energy Physics, Modeling, and Simulations, Faculty of Sciences, University Mohammed V-Agdal, Rabat, Morocco.} 
\maketitle

\section{Introduction}
\label{section:my0}
\indent 
Quantum computing is an exciting and quickly growing field that could change many areas of science and technology. Machine learning (ML) is one of the most promising quantum computing applications, where quantum algorithms can potentially provide exponential speedups over classical algorithms. This field is known as Quantum machine learning (QML).
Quantum machine learning is an emerging field of research that combines the principles of quantum computing and machine learning. QML algorithms can solve complex problems more efficiently and cost-effectively than classical machine learning algorithms. One of the most promising QML algorithms is the quantum support vector machine (QSVM), an extension of the classical support vector machine (SVM) to the quantum realm.
\\
\indent
The classical SVMs are a powerful class of ML algorithms for classification and regression analysis. The development of SVMs can be traced back to the early $1960$s when Vladimir Vapnik and his colleagues began working on a new approach to pattern recognition \cite{svm0, svm1}. However, only in the $1990$s did SVMs gain widespread attention in the ML community \cite{svm2}, thanks to Corinna Cortes and Vladimir Vapnik's pioneering work at AT\&T Bell Labs. They introduced the idea of maximum-margin hyperplanes, decision boundaries that separate data points from different classes with the most significant possible margin \cite{svm3}. 
\\
\indent
This approach allowed SVMs to perform excellent generalization, even with small training datasets. Since then, SVMs have become one of the most extensively used and popular machine learning models and have been successfully applied to various fields, including image recognition, text classification, and bioinformatics. However, as the size of the dataset increases, the computational complexity of SVM also increases, making it difficult to handle large datasets. Still, the QSVM aims to overcome this limitation by leveraging the principles of quantum computing to accelerate the SVM algorithm.
Over the years, there has been significant research in the field of QSVM, exploring various theoretical and practical aspects of the algorithm. 
Researchers have developed several techniques to enhance the performance of QSVM, including the development of quantum kernel methods, quantum feature maps, and quantum optimization techniques.
\\
\indent
One of the early works in QSVM was proposed by Rebentrost \textit{et al.} in $2014$ \cite{A1}, which introduced a quantum algorithm for SVM classification that provides an exponential speedup over classical algorithms. Another essential aspect of QSVM is its robustness to noise. 
In $2015$, Li \textit{et al.} demonstrated a QML algorithm for handwriting recognition on a four-qubit nuclear magnetic resonance (NMR) test bench \cite{liz}; the authors argued that quantum speedup would be highly attractive for tackling significant data challenges. However, this algorithm was specific to NMR-based systems and could not be easily applied to other QML platforms. 
\\
\indent
And after different interesting works, in $2019$, Havlicek \textit{et al.} demonstrated that supervised quantum machine learning models, including QSVM \cite{A2}, can be robust to noise, increasing their practicality for real-world applications.
Subsequently, several studies have been conducted to improve the performance of QSVM, including using quantum feature maps for kernel-based learning, as proposed by Park \textit{et al.} in $2020$ \cite{A3}. Another interesting research explores the potential use of quantum state encoding as a nonlinear feature map, enabling efficient computations in a large Hilbert space efficiently and proposes two approaches for building a quantum model for classification, illustrated with mini-benchmark datasets \cite{A4}.
\\
\indent
In contrast, Liu \textit{et al.} established a rigorous quantum speedup for supervised classification using a general-purpose quantum learning algorithm that only requires classical access to data. This algorithm represents a significant advancement \cite{liu}. In the same year, Schuld \textit{et al.} investigated the impact of quantum data encoding strategies on the efficacy of parametrized quantum circuits in approximating functions \cite{sweke}, and the authors showed that quantum models could realize all possible sets of Fourier coefficients. Therefore, if the accessible frequency spectrum is asymptotically rich enough, such models are universal function approximators. This result has significant implications for developing QML algorithms to tackle complex data challenges.
\\
\indent
In another $2021$ paper \cite{superv}, Schuld, M explored the theoretical foundations of the link between quantum computing and kernel methods in machine learning, systematically rephrased supervised QML models as kernel methods, replacing many near-term and fault-tolerant QML models with a general SVM whose kernel computes distances between data-encoding quantum states. This approach has the potential to significantly reduce the complexity of QML algorithms and improve their performance.
\\
\indent In $2022$, Zhang \textit{et al.} proposed a new quantum optimization algorithm for QSVM that can improve the efficiency and scalability of QSVM on large datasets \cite{A5}. These advancements in QSVM and its related techniques have made it a promising candidate for solving complex problems in various fields, including bioinformatics, finance, image recognition, and material physics.
One of the recent works in QSVM was proposed by Jiang \textit{et al.} in $2023$ \cite{A6}, which introduced a quantum algorithm for SVM classification that leverages the quantum phase estimation algorithm to estimate the kernel matrix. This approach leads to significant speedup compared to classical SVM algorithms, making QSVM a more efficient choice for large-scale datasets.
\\
\indent
In this paper, we build upon these studies and suggest a new QML model for classification that merges the two more accurate approaches identified by Schuld, M and the different works mentioned above. Our model leverages the expressive power of parametrized quantum circuits as function approximators and uses a kernel method to compute distances between data-encoding quantum states. We present theoretical analyses and numerical simulations to demonstrate the potential of our model for tackling classification tasks. Our results suggest that our model outperforms existing QML algorithms, highlighting its potential for future real-world problems and applications.
\\
This paper is divided as follows: \\
 \indent In \S\ref{section:my}, We provide an overview of classical support vector machines, highlighting their principal features and limitations that motivate the exploration of more accurate implementations in quantum machine learning. \\
\indent In \S\ref{section:my1}, we describe the quantum model for support vector machines and explain the three implementations, including our proposed approach, the quantum variational kernel SVM. \\
\indent In \S\ref{section:my2}, we present the results obtained using Pennylane by comparing the accuracy, loss, and confusion matrix indicators of the three quantum SVM models on the Iris dataset. \\
\indent In \S\ref{section:my3}, we discuss our findings' implications and highlight future research directions in this field.
\section{Classical Support Vector Machine}
\label{section:my}

\begin{figure}[!h]
     \centering
     \begin{subfigure}[b]{0.5\textwidth}
         \centering
         \includegraphics[width=\textwidth]{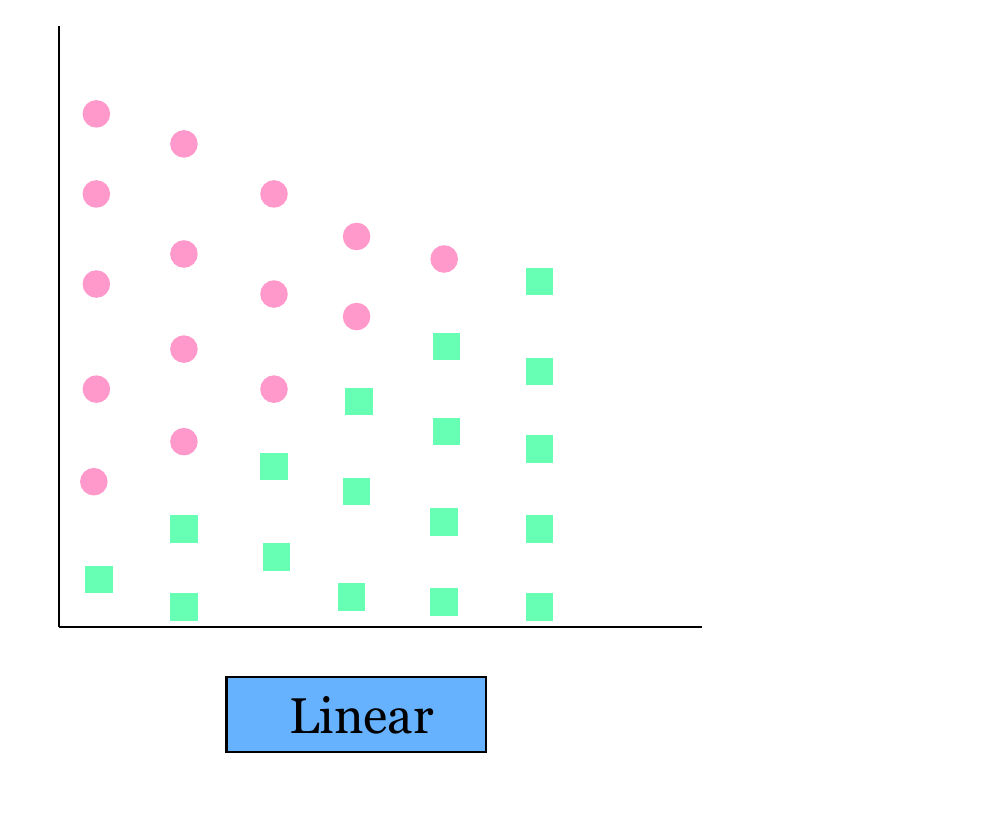}
         \caption{A linear problem}
         \label{s1}
     \end{subfigure}
     \hfill
     \begin{subfigure}[b]{0.45\textwidth}
         \centering
         \includegraphics[width=\textwidth]{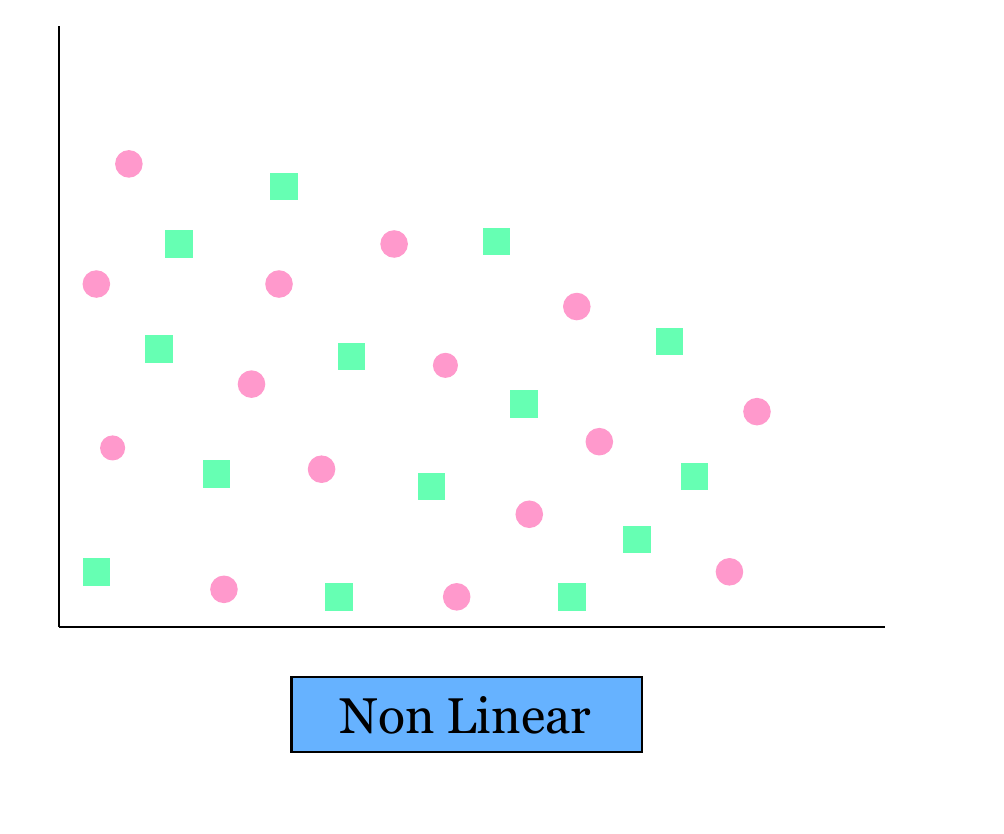}
         \caption{A Non-linear problem}
         \label{ss1}
     \end{subfigure}
        \caption{Graphical representation of linear and non-linear SVMs problems.}
        \label{S}
\end{figure}

In classical machine learning, SVMs are used in supervised models to analyze data for classification and regression. By using this algorithm, we can also perform binary classification and multi-classification. To understand this, we are taking an example. For simplicity, we are taking a binary classification example. Suppose we have a collection of circles and rectangles in a 2D plane. Our job is to classify the circles and rectangles. This problem has two types, (a) linear and (b) non-linear, as shown in Fig.\ref{S}. 

\subsection{Linear SVMs}

\begin{figure}[!h]
    \centering
    \includegraphics[scale=0.8]{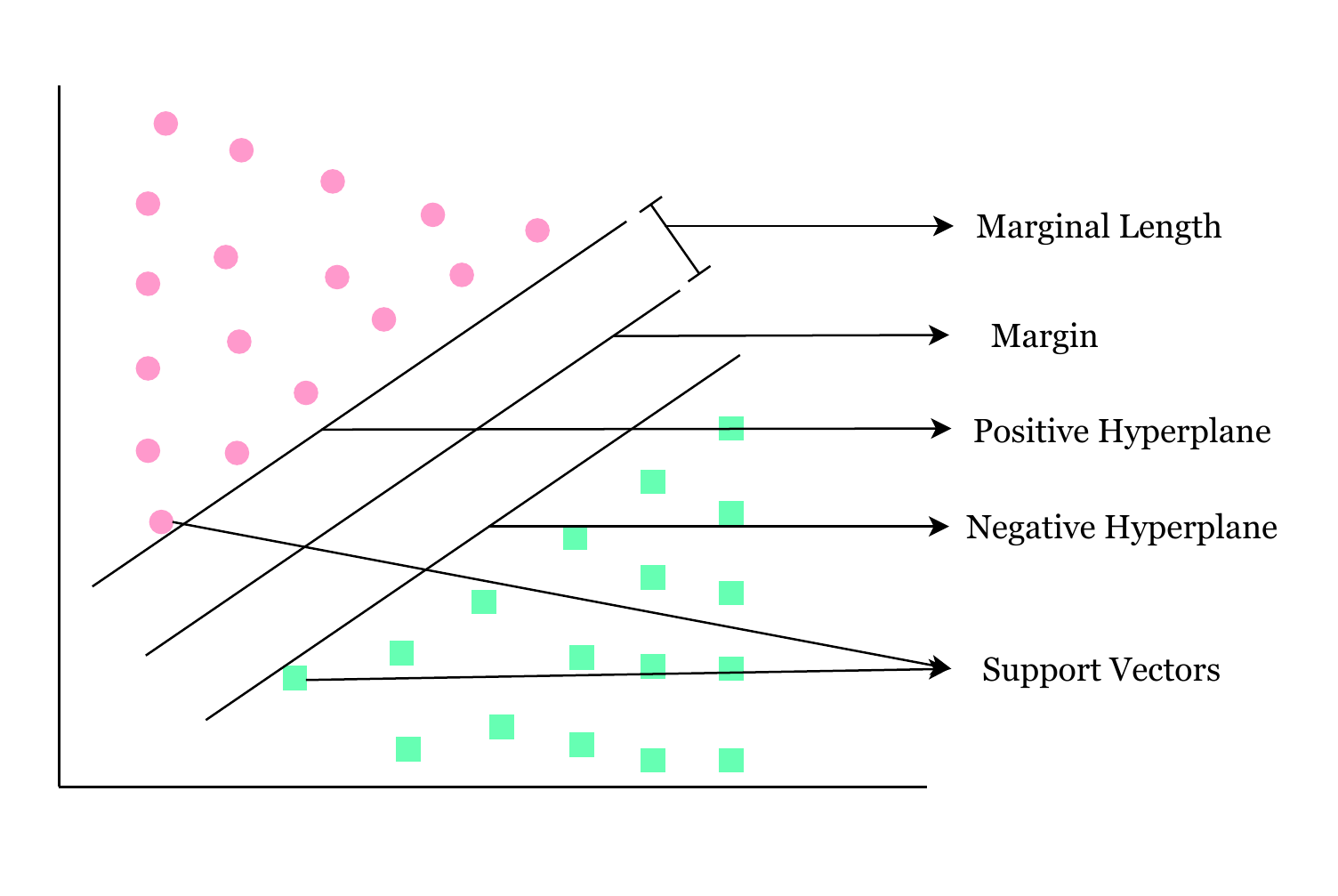}
    \caption{Geometric components of support vector machines.}
    \label{s2}
\end{figure}

First of all, we are discussing linear SVMs. We take a dataset of $n$ points of the form $(x_{1}, y_{1}), (x_{2}, y_{2}), ... (x_{n}, y_{n})$. Here $y_{i}$ are either $1$ or $-1$, and each $x_{i}$ is a p-dimensional real vector. We have to draw the positive hyperplane $(H_{+})$, negative hyperplane $(H_{-})$, and margin as shown in Fig.\ref{s2}. We can find the margin using the formula $= H_{+} + H_{-}$.

Given a $D$-dimensional vector $\textbf{X}_{0}\in\mathbb{R}^{D\times 1}$, and a $(D-1)$-dimensional linear hyperplane $\mathcal{H}:\textbf{W}^{T}\textbf{X}+\textbf{B}-\mathbf{Y}=\boldsymbol{0}$, where $\mathbf{W}=\left(w_{1},w_{2},\ldots,w_{n}\right)$ is the weights vector, $\mathbf{B}$ is the bias vector, and $\Phi(\mathbf{X}_{n})$ is the projection of the point $\mathbf{X}_{n}$ into the nonlinear feature space. The goal is to ascertain the hyperplane that optimally separates the vectorial points into classes while maximizing the margin between the hyperplane and the closest datapoints from each class. Mathematically, we translate this as a quadratic programming problem with linear constraints, whereby our goal is to determine the value of the weights that will maximize the margin
\begin{equation}
\mathbf{W}^{*}=\underset{\mathbf{W}}{\text{arg max}}\;\frac{1}{||\mathbf{W}||_{2}}\left\{\underset{n}{\min}\;\mathbf{Y}_{n}\left[\mathbf{W}^{T}\Phi(\mathbf{X}_{n})+\mathbf{B}\right]\right\},
\end{equation}
where $||\mathbf{W}||_{2}=\left(\sum_{i=1}^{n}w_{i}^{2}\right)^{1/2}$. Mathematically, we can translate this into the primal form SVM optimization problem
$
 \underset{n}{\min}\;\frac{1}{2}||\mathbf{W}||_{2}\quad\text{subject to}\;\mathbf{Y}_{n}\left[\mathbf{W}^{T}\Phi(\mathbf{X}_{n})+\mathbf{B}\right]\geq \mathbf{1}\;\text{for every}\;n. 
$

\subsection{Non-linear SVMs}
As shown in Fig.\ref{s2}, the support vectors are the vectors utilized to generate both the positive and negative hyperplanes. Maximizing the margin length in this specific model is imperative to achieve precise classification and high accuracy.

In order to effectively tackle the non-linear problem we are facing, the kernel trick presents a compelling solution. This technique involves using a kernel function with data points, acquiring higher-dimensional vector points in our feature space. A plethora of kernel functions exists, each tailored to solve different problems. Below, we present a comprehensive list of some of these kernel functions:
\begin{itemize}
\item Polynomial (Homogeneous):
denoted as $K(a_i, a_j) = (a_i \cdot a_j)^d$, where $d$ is a positive integer that determines the degree of the polynomial. By setting $d$ to $1$, it becomes a linear kernel that is particularly useful for linearly separable data.

\item Polynomial (Inhomogeneous): which incorporates a constant term $r$ to the dot product of the input vectors, resulting in $K(a_i, a_j) = (a_i \cdot a_j + r)^d$. This kernel is well-suited for capturing nonlinear relationships between the data.

\item Sigmoid function (Hyperbolic tangent): based on the hyperbolic tangent function, takes the form $K(a_i, a_j) = \tanh(ka_i \cdot a_j + c )^d$, where $k$ and $c$ are kernel parameters. This kernel can be used to model data that exhibits sigmoidal behavior and has been applied in various applications such as image classification and text mining.
\end{itemize}

\indent After applying the kernel function to our data points, we have to do the same operation as in linear. Then we can complete the classification successfully.
We modify the primal form of the linear SVM to include the slack variables $\boldsymbol{\xi}\geq \boldsymbol{0}$
$
\underset{n}{\min}\;\frac{1}{2}||\mathbf{W}||_{2}+\mathbf{C}\sum_{n}\boldsymbol{\xi}_{n}\quad\text{subject to}\;\mathbf{Y}_{n}\left[\mathbf{W}^{T}\Phi(\mathbf{X}_{n})+\mathbf{B}\right]\geq 1-\boldsymbol{\xi}_{n}\;\text{for every}\;n.
$
In addition to quadratic programming, numerous alternative techniques exist for resolving this problem. These include the approaches of Lagrange multipliers, sequential minimal optimization, interior point methods, gradient descent (GD), stochastic gradient descent (SGD), and kernel methods.


\subsection{Disadvantages of Classical SVMs}
Despite the popularity of classical SVMs, they have certain limitations that constrain their optimal performance. One of the significant limitations is handling high-dimensional feature spaces, which can result in slow training times and overfitting problems. Another area for improvement is the dependence on kernel functions, which may not effectively capture complex data relationships. Furthermore, classical SVMs are not easily scalable to large datasets and demand extensive parameter tuning for accurate results. Researchers have turned to quantum machine learning to overcome these limitations and explore more precise and efficient alternatives. In the next section, we examine how quantum support vector machines can effectively tackle these challenges and provide a promising solution for enhancing classification performance.

\section{Quantum Support Vector Machine}
\label{section:my1}
Quantum Support Vector Machine is a burgeoning area of research in quantum machine learning that offers promising potential for enhanced computational performance in classification and regression tasks. While classical SVM has been widely utilized in machine learning, QSVM exploits the unique properties of quantum mechanics to outperform classical SVM in specific applications. The QSVM algorithm involves mapping input data onto a quantum state, which is subsequently subjected to quantum circuit processing to generate the classification outcome. The described circuit comprises a sequence of quantum gates that manipulate the quantum state and execute the SVM algorithm. The classification outcome is obtained by measuring the circuit's output.
\\
\indent In previous works, as mentioned in the introduction sections (\ref{section:my0}), QSVMs have been implemented using various approaches, such as the quantum kernel method, the quantum matrix inversion method, and the quantum feature mapping method. Nevertheless, these methodologies possess certain constraints, such as high error rates, enormous computational resources, and scalability issues. This section will focus on three recent approaches to QSVM: the quantum kernel approach, the quantum variational approach, and a novel hybrid approach that combines the quantum kernel approach with the quantum variational circuit. These approaches have shown promising results and offer potential accuracy, scalability, and robustness improvements. The following subsections will describe each approach, highlighting the steps and circuits used to develop the QSVM models.

\begin{figure}[!h]
    \centering
    \includegraphics[scale=1]{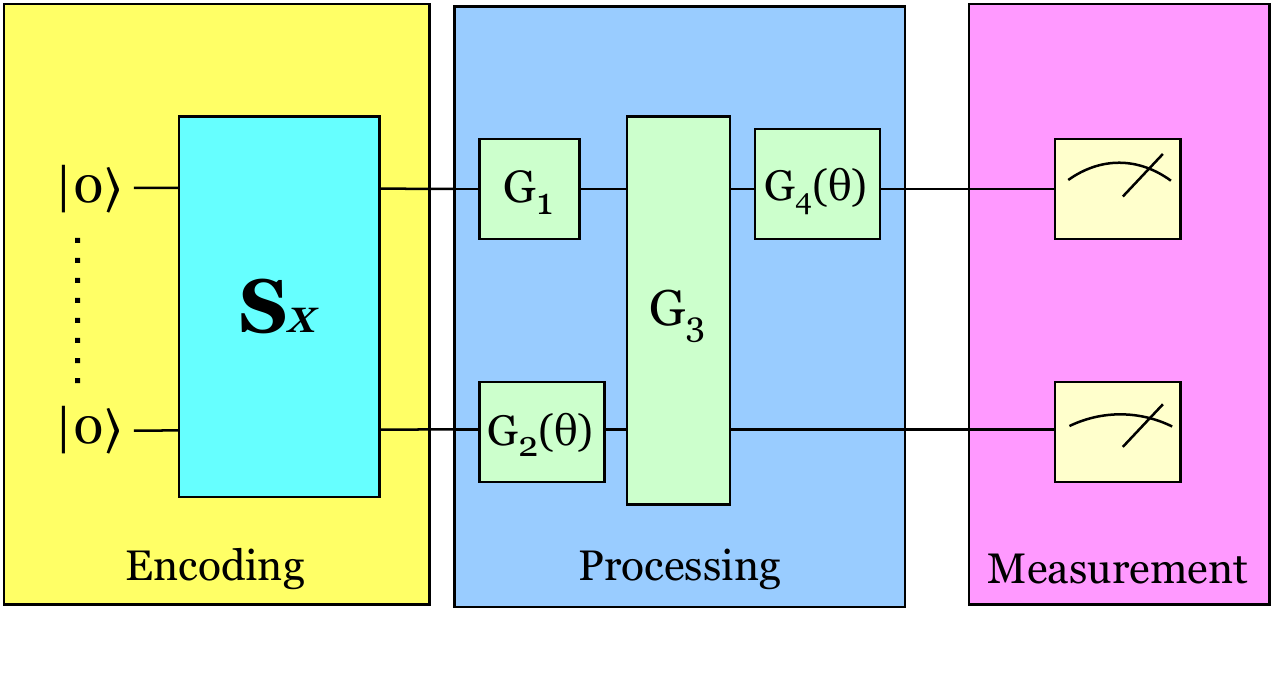}
    \caption{Quantum circuit architecture for QSVMS: Generalized model description.}
    \label{c2}
\end{figure}

In each approach, the first step in our methodology involves the conversion of our classical datapoints into quantum states. To achieve this, we begin by encoding the data points using a quantum circuit, as depicted in Fig.\ref{c2}. Subsequently, we establish our data set and opt for the iris data set for simplicity. Our selection of qubits is based on the features outlined in our data set. We utilize a quantum model represented as follows:
\begin{equation}
f(x) = \langle \phi(x)| M | \phi(x)\rangle.
\end{equation}
Here $|\phi(x)\rangle$ is prepared using an encoding circuit, and $M$ is the measurement operator. $M$ is observable that is defined as:
\begin{equation}
    M(\theta) = G^{\dagger}(\theta)\sigma^{0}_z G(\theta).
\end{equation}

\subsection{Quantum Kernel Support Vector Machine}
\label{subsection:qk}

\begin{figure}[!h]
    \centering
    \includegraphics[scale=0.8]{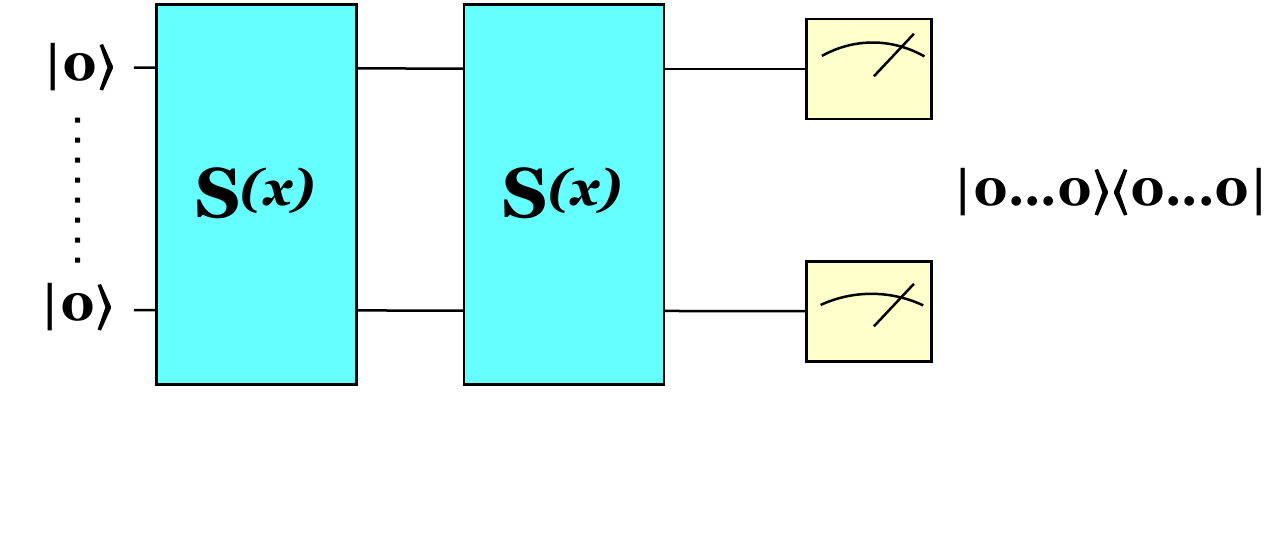}
    \caption{QK-SVM circuit.}
    \label{c1}
\end{figure}

\begin{figure}[!h]
    \centering
    \includegraphics[scale=0.6]{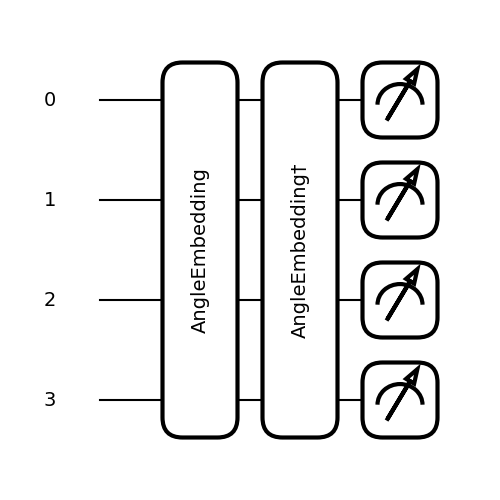}
    \caption{QK-SVM circuit using Pennlyne.}
    \label{qc2}
\end{figure}

We propose implementing support vector machines with a kernel computed by a quantum circuit in this approach. Specifically, we utilize the angle-embedding template in conjunction with a SWAP test to evaluate the quantum kernel; this method reduces the number of required qubits by half, rendering it more viable for real-world implementations.
\\
\indent
The kernel function is a central concept in SVMs, which are a popular class of ML algorithms for classification and regression tasks; this kernel function is a measure of similarity between two data points in a high-dimensional feature space and is used to map the data into a space where a linear classifier can separate the classes. This method can be used with any kernel function, like the linear kernel and radial basis function (RBF) kernel, computed using a quantum circuit, and we call it the quantum kernel.
Mathematically, the quantum kernel is represented by the following equation:
\begin{equation}
k(x_1, x_2) = |\langle \phi(x_1)|\phi(x_2)\rangle|^2,
\end{equation}
Where $x_1$ and $x_2$ are the input feature vectors, and $\phi(x_i)_{i=1, 2}$ denotes the quantum embedding of $x$ into a quantum state with the angle encoding routines $S(x_1)$ and $S(x_2)$, we then apply the inverse embedding to one of the states and compute the overlap between the two states using a SWAP test, and the SWAP test is a simple quantum protocol that measures the overlap between two quantum states, we can represent this step by the following equation:
\begin{equation}
    \langle SWAP \rangle = |\langle \phi(x_1) \otimes \phi(x_2) | SWAP | \phi(x_1) \otimes \phi(x_2) \rangle|^2,
\end{equation}
$SWAP$ is the swap gate, and $|\langle SWAP \rangle|^2$ represents the probability of measuring the two quantum embeddings in the same state. Finally, we use the Hermitian observable to measure the projector onto the initial state $|0...0\rangle\langle 0...0|$, and Fig.\ref{c1} present this circuit.
\\
\indent The advantage of this approach is that it has the potential to scale to larger datasets by utilizing quantum hardware with more qubits. As we mentioned, it also requires only half the number of qubits as the number of features, and this is because we can prepare the two data points on the same set of qubits using the angle-embedding template and then apply the inverse embedding to one of the states, as shown in Fig.\ref{qc2} using Pennylane \cite{kr}.


\subsection{Quantum Variational Support Vector Machine}
\label{subsection:qv}
In this method, we propose a novel approach for training data directly by utilizing an ansatz for the variational circuit. This ansatz, a quantum operation applied in multiple layers, enhances expressivity. Although the variational circuit cannot optimize the exact cost, similar to SVMs, we have incorporated a bias term termed hinge loss in our quantum model to minimize the gap between the two. And in the quantum node, we explicitly apply the parameter shift differentiation method. The variational quantum circuit is given in Fig.\ref{c2}. We have given our method's encoding, processing, and measurement steps in this circuit. The quantum variational method is a key concept in quantum machine learning, a rapidly growing field that aims to leverage quantum computing to develop robust machine learning algorithms. This answer will discuss the quantum variational method and its applications in quantum machine learning. 

Mathematically, the quantum variational method can be described as follows:\\
Suppose we have a parameterized quantum circuit that can be represented by the unitary operator $U(\theta)$, where $\theta$ is a vector of parameters. Given a set of training data {(x$_1$, y$_1$), (x$_2$, y$_2$), ..., (x$_n$, y$_n$)}, where x$_i$ is an input and y$_i$ is the desired output. We want to find the values of $\theta$ that minimize the cost function:
\begin{equation}
    f(\theta) = \frac{1}{n} \sum_{i=1}^n L(y_i, U(\theta)x_i).
\end{equation}

Here, $L(y, y')$ measures the difference between the desired output y and the actual output y' produced by the quantum circuit. This cost function is typically chosen to be a function that can be efficiently computed on a classical computer. We use an iterative optimization algorithm such as gradient descent to find the optimal values of $\theta$ that minimize the cost function. Starting from an initial guess for $\theta$, we compute the gradient of the cost function concerning each parameter and update the parameters in the direction of the negative gradient. This process is repeated until the cost function converges to a minimum. The circuit structure is given below in Fig.\ref{qc1}.

\begin{figure}[!h]
    \centering
    \includegraphics[scale=0.6]{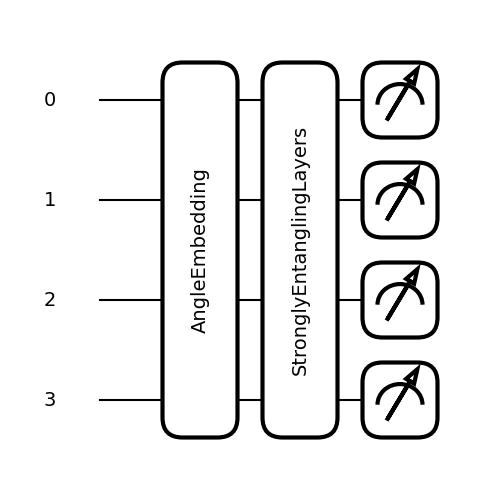}
    \caption{QV-SVM circuit using Pennlyne.}
    \label{qc1}
\end{figure}

\subsection{Quantum Variational Kernel Support Vector Machine}
\label{subsection:qvk}
This study proposes a new approach for quantum support vector machines, which we call Quantum Variational Kernel Support Vector Machine (QVK-SVM). It combines two distinct methods to enhance the performance of quantum kernels and variational circuits. The first method utilizes the angle-embedding template to prepare the quantum states used to compute the kernel, as we explained in subsection \ref{subsection:qk}. The overlap between the two states is measured using a SWAP test, which requires only half the qubits. 

The second method involves utilizing a variational circuit trained through the variational training principle, as outlined in Subsection \ref{subsection:qv}. The ansatz of the circuit can be improved by adding more layers, thereby enhancing its ability to express itself. Additionally, a bias term is incorporated into the quantum model to facilitate training on the hinge loss. The quantum node utilizes the parameter-shift differentiation method, which is very efficient on hardware.

\begin{figure}[!h]
    \centering
    \includegraphics[scale=0.6]{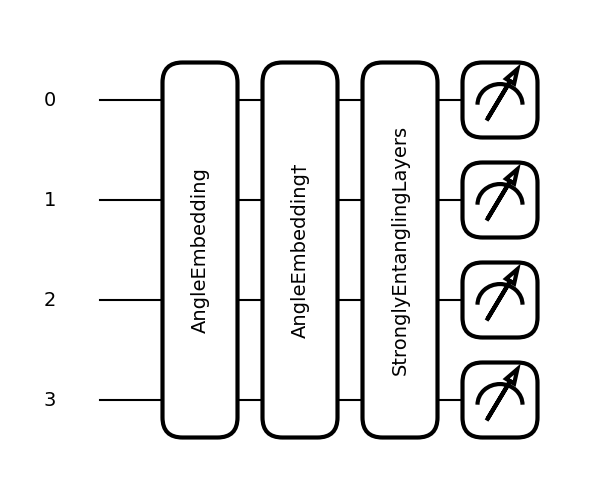}
    \caption{QVK-SVM circuit using Pennlyne.}
    \label{qcK}
\end{figure}

The proposed circuit of the new approach consists of three main components: AngleEmbedding, the adjoint of AngleEmbedding, and StronglyEntanglingLayers, as shown in Fig.\ref{qcK}.
 
The AngleEmbedding is used to prepare the quantum states of the data points, which are then fed into the adjoint of the AngleEmbedding to prepare the inverse embedding. The StronglyEntanglingLayers component is used to apply the variational circuit, which is trained using the hinge loss. 

The proposed approach has several advantages. First, it combines the strengths of both methods to enhance the performance of QSVMs. Second, it utilizes the variational training principle, allowing greater control over the training process. Third, it uses the parameter-shift differentiation method, which works well on hardware. Finally, the proposed circuit is simple and easy to implement, making it suitable for practical applications.

 The proposed approach for quantum SVMs combines the angle-embedding kernel and the variational circuit to enhance the performance of QSVMs. The proposed approach has several advantages over the existing methods, including greater control over the training process, better hardware compatibility, and ease of implementation. Future research could explore the application of the proposed approach to other datasets and investigate the potential of the approach for solving more complex problems.
\section{Results and Discussion}
\label{section:my2}
The results and discussion of the three models in this research demonstrate the potential of quantum machine learning in enhancing binary classification tasks, particularly the quantum support vector machine. The first model, QK-SVM, employed a quantum kernel approach and delivered an impressive overall accuracy of $96.34\%$ on the test set. The second model, QV-SVM, utilized variational training with a quantum support vector machine and achieved a maximum accuracy of $95.43\%$ on the test set. The third and final model, QVK-SVM, combined quantum kernel and variational training to yield the most promising results, with an accuracy of $98.48\%$ on the test set, as evidenced by the data presented in Table \ref{tab:matrix}.
\begin{table}[htbp]
\centering
\caption{Comparison of confusion matrix indicators for QK-SVM, QV-SVM, and QVK-SVM models.}
\label{tab:matrix}
\begin{tabular}{cccccc}
\hline
\textbf{Model} & \textbf{Accuracy} & \textbf{Precision} & \textbf{Recall} & \textbf{Specificity} & \textbf{F1 Score} \\
\hline
QK-SVM & 96.34\% & 93.77\% & 89.64\% & 95.11\% & 91.64\% 
\\
QV-SVM & 95.43\% & 90.89\% & 89.11\% & 92.88\% & 89.99\%
\\
QVK-SVM & 98.48\% & 96.27\% & 92.35\% & 97.85\% & 94.24\% \\
\hline
\end{tabular}
\end{table}

Table \ref{tab:matrix} displays the performance metrics for the three models: QK-SVM, QV-SVM, and QVK-SVM. 
Our findings indicate that QK-SVM achieved high precision and recall rates, indicating the robustness of the model in correctly identifying the different classes of Iris flowers. QV-SVM achieved high specificity and F1 score values, further validating the effectiveness of the quantum support vector machine using the variational algorithm. QVK-SVM achieved high precision, recall, and specificity; these results confirm the efficacy of QVK-SVM and highlight its potential as a reliable tool for ML applications.

\begin{figure}[htbp]
\hspace{-2cm}
  \begin{minipage}[t]
  {0.5\linewidth}

    \includegraphics[width=1.3\linewidth]{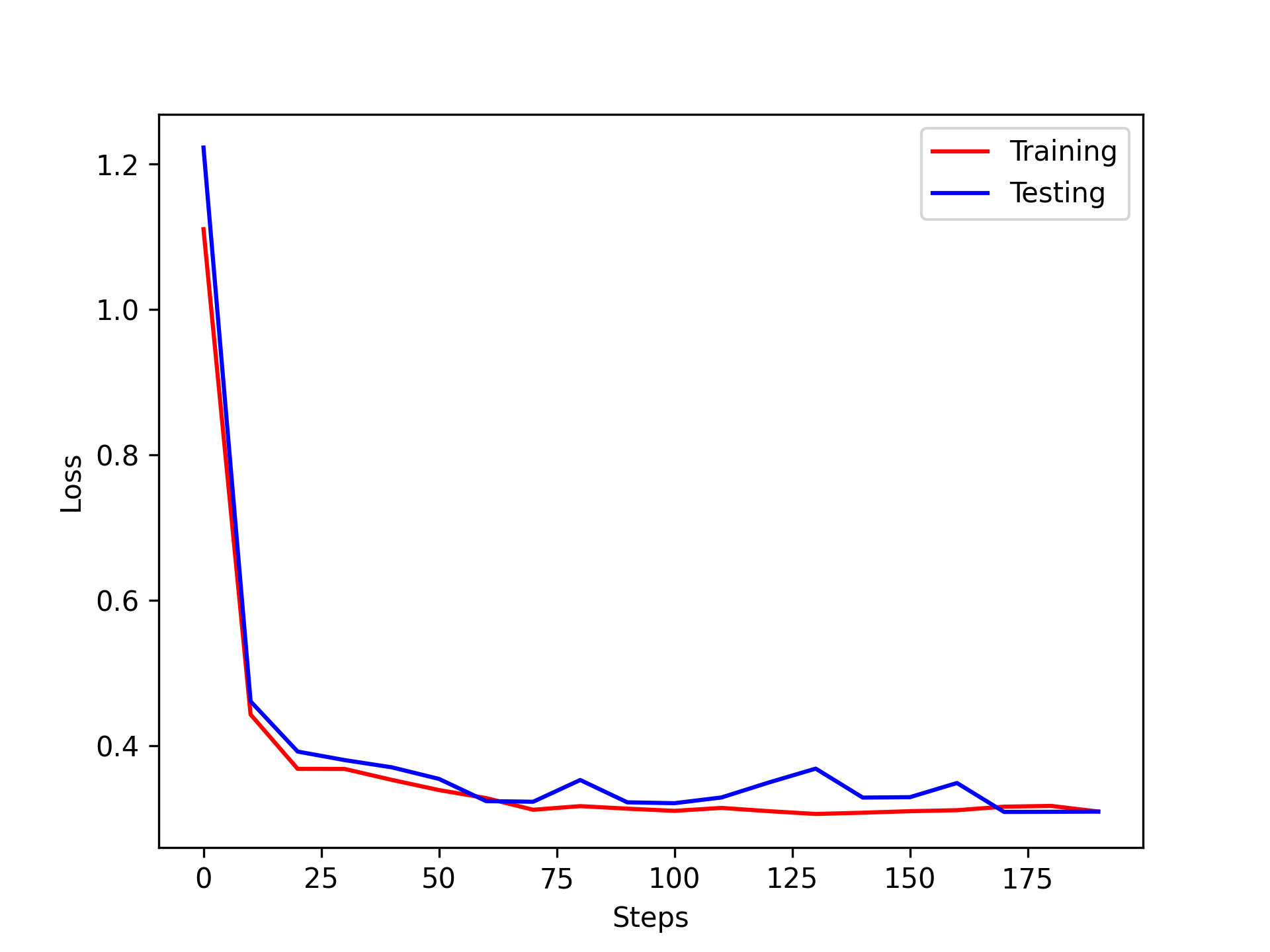}
    \caption{Graph of loss function of the QK-SVM.}
    \label{fig:loss2}
  \end{minipage}\hfill
  \hspace{1cm}
  \begin{minipage}[t]{0.5\linewidth}

    \includegraphics[width=1.3\linewidth]{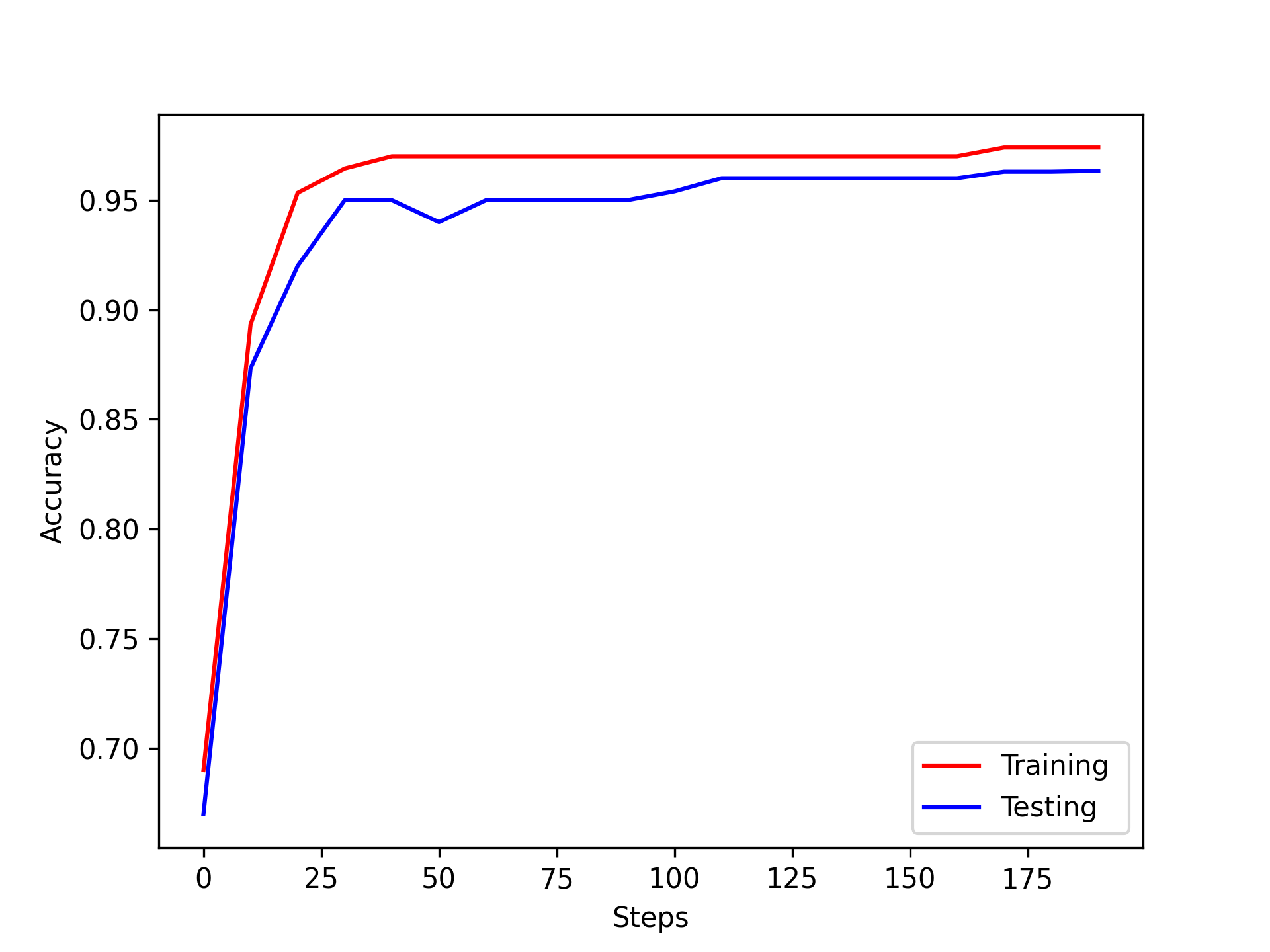}
    \caption{Accuracy line graph of the QK-SVM.}
    \label{fig:acc2}
  \end{minipage}
\end{figure}

The QK-SVM model achieved a maximum accuracy of $96.34\%$, high precision and recall rates, and a corresponding specificity and F1 score. Fig.\ref{fig:loss2} and Fig.\ref{fig:acc2} illustrate the loss and accuracy curves for both the training and testing sets, demonstrating a clear improvement in the model's performance during training.

\begin{figure}[htbp]
\hspace{-2cm}
  \begin{minipage}[t]
  {0.5\linewidth}

    \includegraphics[width=1.3\linewidth]{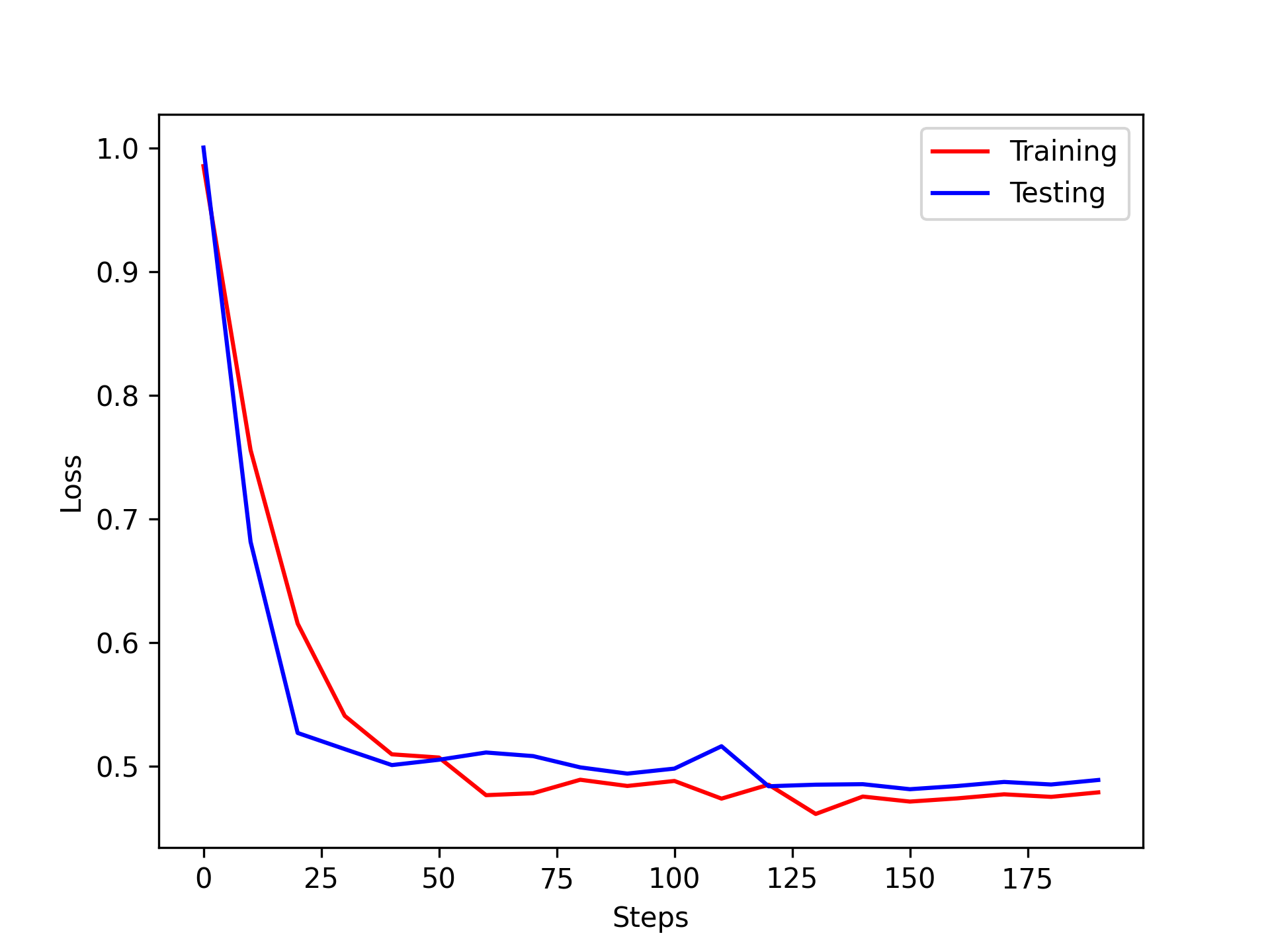}
    \caption{Graph of loss function of the QV-SVM.}
    \label{fig:loss1}
  \end{minipage}\hfill
  \hspace{1cm}
  \begin{minipage}[t]{0.5\linewidth}

    \includegraphics[width=1.3\linewidth]{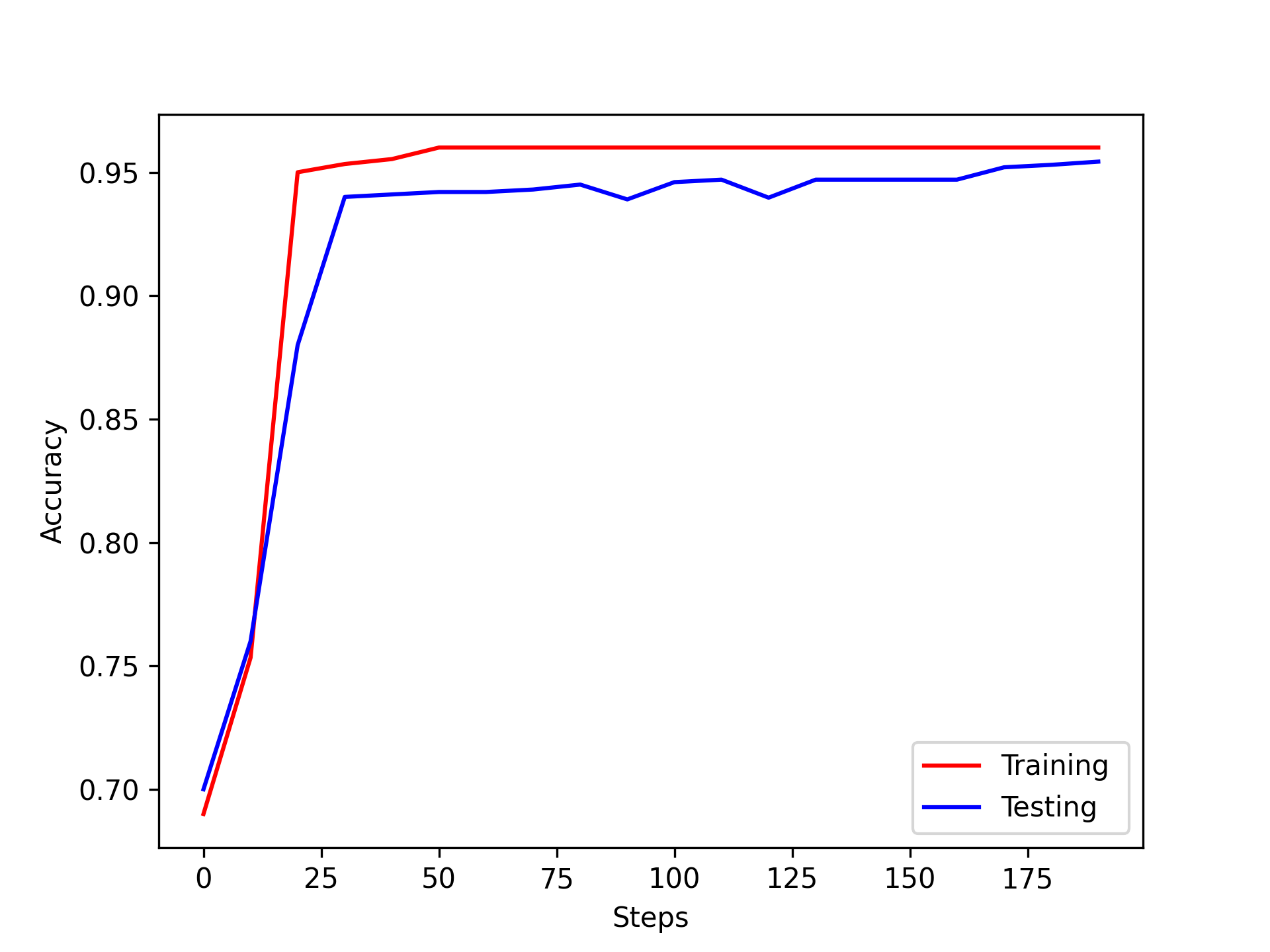}
    \caption{Accuracy line graph of the QV-SVM.}
    \label{fig:acc1}
  \end{minipage}
\end{figure}

The QV-SVM model achieved a steady improvement in performance as the number of iterations increased, with training losses ranging from $1.00$ to $0.49$ and testing losses ranging from $0.98$ to $0.47$. The model achieved an absolute accuracy of $95.43\%$, with high precision and recall rates. Fig.\ref{fig:loss1} and Fig.\ref{fig:acc1} illustrate the loss and accuracy plots, respectively, highlighting the model's optimization over the iterations.

Our suggested model represents a novel approach that combines the quantum kernel and variational algorithm used in QV-SVM and QK-SVM, respectively. The results of our QVK-SVM model indicate that this combined approach is highly effective for solving classification problems, even on relatively small datasets.

\begin{figure}[htbp]
\hspace{-2cm}
  \begin{minipage}[t]
  {0.5\linewidth}

    \includegraphics[width=1.3\linewidth]{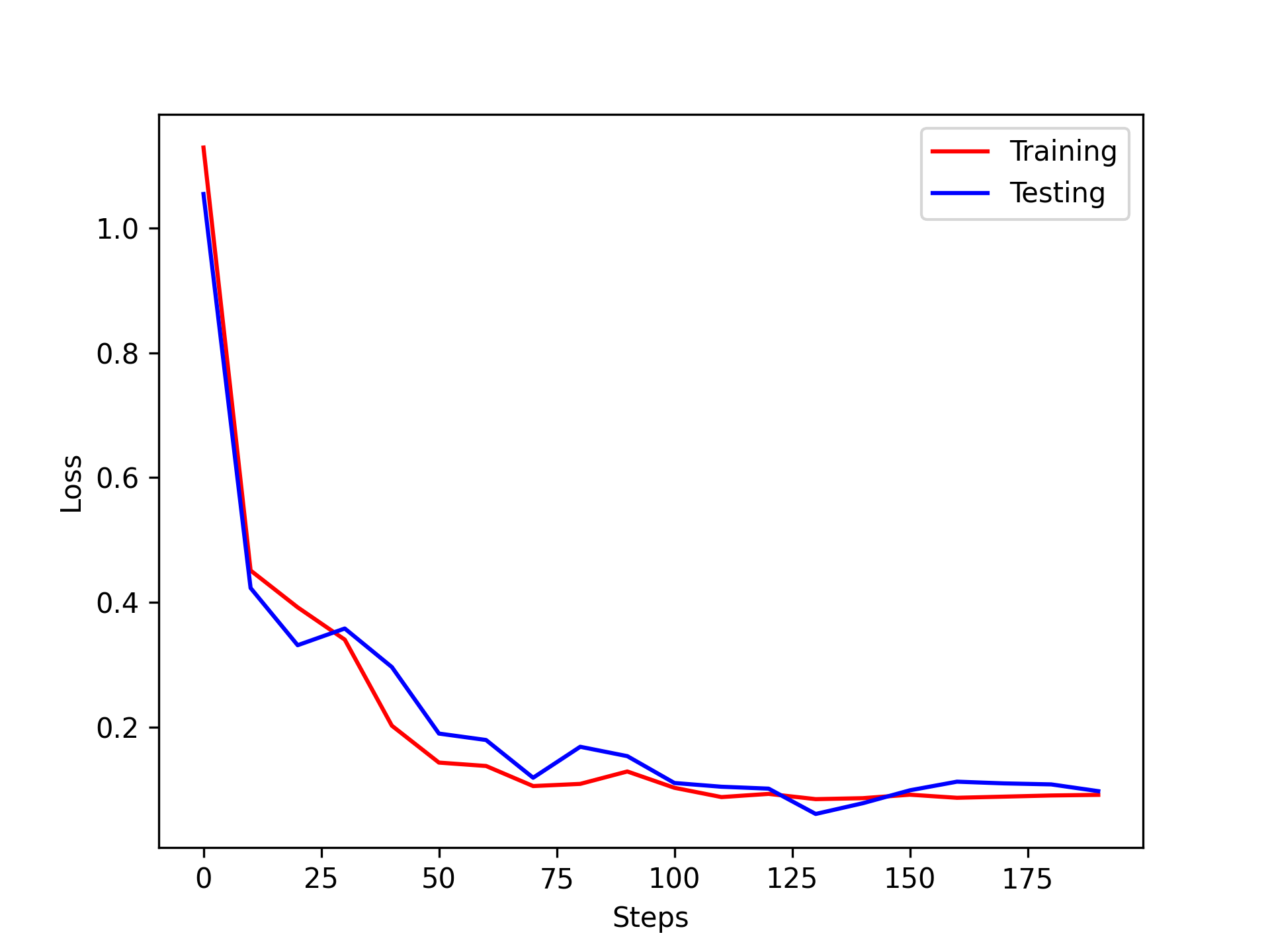}
    \caption{Graph of loss function of the QVK-SVM.}
    \label{fig:loss3}
  \end{minipage}\hfill
  \hspace{1cm}
  \begin{minipage}[t]{0.5\linewidth}

    \includegraphics[width=1.3\linewidth]{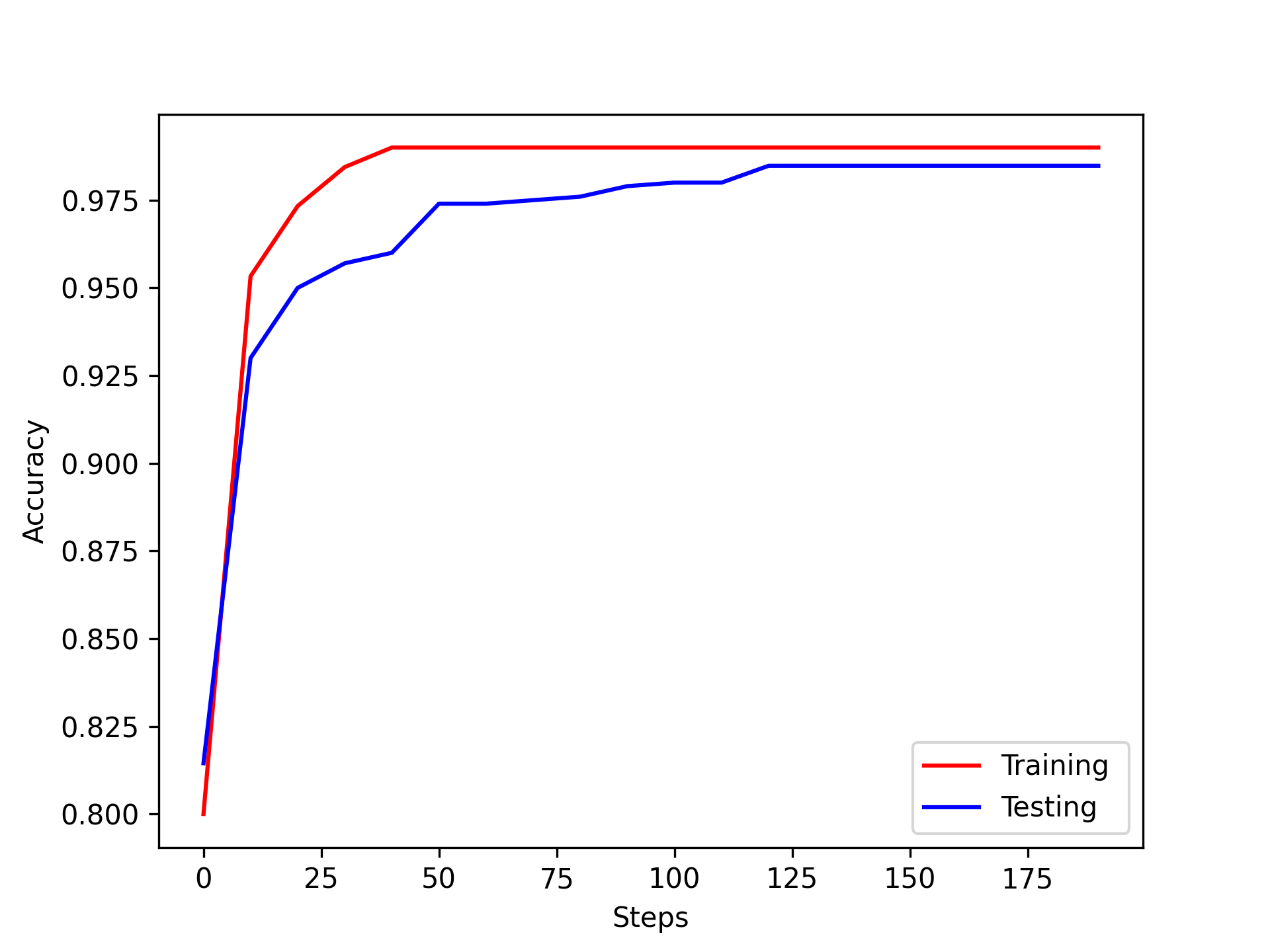}
    \caption{Accuracy line graph of the QVK-SVM.}
    \label{fig:acc3}
  \end{minipage}
\end{figure}

The proposed QVK-SVM model achieved an impressive accuracy of $98.48\%$ on the test set, with a corresponding F1 score of $91.64\%$. Fig.\ref{fig:loss3} shows the convergence of the training and testing losses throughout the experiment, demonstrating that the model could optimize the loss function to achieve high accuracy. Fig.\ref{fig:acc3} displays the corresponding training and testing accuracies, showing that the model's accuracy improved steadily throughout the experiment, reaching a peak of $98\%$ accuracy on the test set. 

Furthermore, the superior performance of QVK-SVM, compared to QV-SVM and QK-SVM, suggests that the combination of these approaches can provide a more robust and reliable solution to binary classification tasks.

The results show the potential of quantum machine learning in enhancing binary classification tasks, especially the quantum support vector machine. As demonstrated in our novel method, combining the quantum kernel and variational algorithm represents a promising approach that could be extended to other datasets and classification problems.

\section{Conclusion}
\label{section:my3}
This study delved into applying quantum support vector machines for binary classification tasks. Specifically, two pre-existing methods, the quantum kernel support vector machine and the quantum variational support vector machine, were compared and evaluated for their respective strengths and limitations. In addition, a novel approach was developed by combining these two methods, resulting in the quantum variational kernel support vector machine. The proposed QVK-SVM approach demonstrated exceptional accuracy and loss reduction performance, offering a promising solution for binary classification tasks in quantum machine learning. These findings hold significant potential for advancing the field of quantum machine learning and its diverse applications. The QVK-SVM approach represents a noteworthy contribution to this field's development and has clear implications for future research endeavors.
\\ 
\indent The proposed method presents an opportunity for further research to explore its efficacy in resolving intricate issues across various datasets. Advanced optimization techniques and the development of new quantum algorithms can enhance the efficiency and scalability of the approach. Furthermore, the potential of quantum machine learning can be investigated by extending the proposed method to other machine learning models, such as neural networks and decision trees. Through these efforts, the proposed approach can advance the field of quantum machine learning and unlock new opportunities for addressing complex real-world problems. Its potential to do so is significant and warrants further academic investigation.

\end{document}